\documentclass[journal]{IEEEtran}

\usepackage{xcolor,soul,framed} 

\colorlet{shadecolor}{yellow}
\usepackage[pdftex]{graphicx}
\graphicspath{{../pdf/}{../jpeg/}}
\DeclareGraphicsExtensions{.pdf,.jpeg,.png}

\usepackage[cmex10]{amsmath}
\usepackage{array}
\usepackage{mdwmath}
\usepackage{mdwtab}
\usepackage{eqparbox}
\usepackage{url}
\usepackage{float}
\usepackage{authblk}
\usepackage[square,sort,compress,comma,numbers]{natbib}
\usepackage{algpseudocode}
\usepackage{hyperref}
\usepackage{graphicx}
\usepackage{subcaption}
\hypersetup{
  colorlinks=true,
  linkcolor=blue,
  filecolor=magenta,      
  urlcolor=cyan,
  citecolor=blue,
  pdftitle={Your Title},
  pdfpagemode=UseOutlines
}

\begin{document}
\bstctlcite{IEEEexample:BSTcontrol}
    \title{A Comparison of Two Lattice Boltzmann Models for Electrodynamics}
        \author[1]{Jorge Isaac Rubiano Murcia \thanks{jrubianom@unal.edu.co}}
        \author[1]{Alejandro Mario Salas Estrada \thanks{asalas@unal.edu.co}}
        \author[1]{Jose David Hernandez Ortega\thanks{johernandezo@unal.edu.co }} 
        \affil[1]{National University of Colombia, Faculty of Sciences, Department of Physics, Bogotá, Colombia.}


\maketitle

\begin{abstract}
In recent years, various Lattice Boltzmann models for electrodynamics have been developed as alternatives to classical methods such as Finite Difference Time Domain (FDTD) and Finite Element Methods (FEM). However, there has been a lack of systematic comparisons between these models. This paper addresses this gap by comparing two specific Lattice Boltzmann models, published by Mendoza and Muñoz (MM), and Hauser and Verhey (HV), respectively.

To compare the models, we utilize time and memory as indicators, considering the same achieved error, in four standard tests: a dielectric pulse traveling through two interfaces, the skin effect, the Hertz dipole, and a dielectric pulse traveling through several interfaces.

The results indicate that both methods accurately simulate the tests and exhibit convergence as the mesh is refined. However, the MM method outperforms the HV method regarding time, while its memory efficiency was lower. The modified Hauser-Verhey model demonstrates itself to be a promising alternative to the Mendoza-Muñoz model. These findings contribute to the ongoing development and optimization of numerical methods for electromagnetics simulations.
\end{abstract}


%
\IEEEpeerreviewmaketitle


\section{Introduction}
The Lattice-Boltzmann method (LBM), although a recent numerical approach compared to other alternatives, has proven to be a powerful tool for simulating processes modeled by conservation equations \cite{Zhang2008}. Among the use cases, we could highlight applications in computational fluid dynamics where LBM allows for the simulation of complex geometries and multi-phase flows \cite{LectureA}. Similarly, it has also been successfully deployed to simulate magnetohydrodynamics \cite{Dellar_2010}, the wave equation \cite{GUANGWU200061}, Poisson's equation \cite{CHAI20082050}, and linear and non-linear Schrödinger equations \cite{QLB}.

More recently, however, it was further extended by Mendoza and Muñoz to integrate the electrodynamic Maxwell's equations \cite{MendozaMunoz}. Following their publication, numerous authors proposed alternative formulations to recover Maxwell's equations. For example, Succi and collaborators \cite{Hanasoge_2011} proposed a scheme to simulate three-dimensional wave propagation in dispersive media; Hauser and Verhey \cite{HV2017,HV2019} similarly treated complex media while significantly reducing the complexity of the scheme, compared with previous approaches. In this same fashion, other alternatives were presented for dispersive media \cite{ieee,CHEN2013961} and electromagnetic waves in one-dimensional photonic crystals \cite{photonics9070464}.

All these publications perform some cornerstone simulations such as point dipole antennas, the Skin effect, and medium changes, and some of them carry out encouraging comparisons with well-established methods such as FDTDM \cite{MendozaMunoz,HV2019}. Nevertheless, there is no systematic comparison between Lattice-Boltzmann formulations for Maxwell's equations in the current literature. What's more concerning, many previous publications suggest some degree of suitability over competing alternatives without solid arguments.

In this work, we fill this literature gap by comparing the schemes' CPU time and memory requirement in \cite{MendozaMunoz} and \cite{HV2017}. For this, we simulated a Gaussian pulse crossing an interface, the Skin effect, the dipole radiation, and a one-dimensional highly non-uniform media, for different refinements of the grid.

\subsection{Bhatnagar Gross and Krook (BGK) scheme}
The Boltzmann equation governs the evolution of an equilibrium function $f$ in the phase space, representing a scalar of the probability distribution of physical quantities in the system. In the Bhatnagar-Gross-Krook (BGK) scheme, the collision operator $\Omega$ is proportional to the deviation of the equilibrium function, driving the system towards thermal equilibrium.

The Boltzmann equation can be expressed as follows, see \cite{Hanasoge_2011,ZhaoliZheng}.

\begin{equation}
\frac{\partial f}{\partial t} + \Vec{v} \cdot \nabla f = \Omega(f) + T,
\end{equation}

where $\Omega(f)$ represents the collision operator and $T$ denotes the source terms that can be included if necessary.

In numerical computations, the velocity set of vectors is discretized into a finite set of directions, while the physical space is discretized into a mesh. The Boltzmann equation can be solved using a two-step process: collision and advection. This can be achieved through the BGK model in the context of the lattice Boltzmann method. The discretized versions of the collision and advection steps can be expressed as follows:

Collision step:
\begin{equation}
f^{*}(\Vec{x}, \Vec{v}, t+\Delta t) = f(\Vec{x}, \Vec{v}, t) - \frac{\Delta t}{\tau}\left(f(\Vec{x}, \Vec{v}, t) - f_{\text{eq}}(\Vec{x}, \Vec{v})\right) + T,
\end{equation}

Advection step:
\begin{equation}
f(\Vec{x} + \Delta t\Vec{v}, \Vec{v}, t + \Delta t) = f^{*}(\Vec{x}, \Vec{v}, t + \Delta t),
\end{equation}
where $\Delta t$ is the time step, $\tau$ is the relaxation time, and $f_{\text{eq}}$ is the equilibrium distribution function. These equations represent the numerical implementation of the BGK scheme within the lattice Boltzmann method.
\subsection{Maxwell's equations}
For linear media, Maxwell's equations can be written in terms of the electric field $\Vec{E}$ and magnetic field $\Vec{B}$, as well as the electric displacement field $\Vec{D}$ and magnetic induction field $\Vec{H}$. The equations are as follows, see \cite{jackson_classical_1999}.
\begin{enumerate}
\item Gauss's law for electric fields:

\begin{equation}
\nabla \cdot \Vec{D} = \rho,
\end{equation}

where $\rho$ is the electric charge density.

\item Gauss's law for magnetic fields:
\begin{equation}
\nabla \cdot \Vec{B} = 0.
\end{equation}

\item Faraday's law of electromagnetic induction:
\begin{equation}
\nabla \times \Vec{E} = -\frac{\partial \Vec{B}}{\partial t}.
\end{equation}

\item Ampere's law with Maxwell's addition:
\begin{equation}
\nabla \times \Vec{H} = \Vec{J} + \frac{\partial \Vec{D}}{\partial t},
\end{equation}
where $\Vec{J}$ is the electric current density.
\end{enumerate}
These equations are supplemented by the continuity equation:
\begin{equation}
\nabla \cdot \Vec{J} = -\frac{\partial \rho}{\partial t},
\end{equation}
which expresses the conservation of charge.

It's important to note that the electric displacement field $\Vec{D}$ is related to the electric field $\Vec{E}$ by:
\begin{equation}
\Vec{D} = \epsilon \Vec{E},
\end{equation}
where $\epsilon$ is the permittivity of the medium.

Similarly, the magnetic induction field $\Vec{H}$ is related to the magnetic field $\Vec{B}$ by:
\begin{equation}
\Vec{H} = \frac{1}{\mu} \Vec{B},
\end{equation}
where $\mu$ is the permeability of the medium.\\
In addition, the Guass's laws can be obtained from the Faradys's law, Ampere-Maxwell law and continuity equation, see \cite{HV2017,MendozaMunoz}.

These equations govern the behavior of electromagnetic fields in linear media.\\
\textit{Remark: Throughout the entire text, the permittivity and permeability are denoted as $\epsilon = \epsilon_r \epsilon_0$ and $\mu = \mu_r \mu_0$, where $\epsilon_0$ and $\mu_0$ are the permittivity and permeability in vacuum. Meanwhile, $\epsilon_r$ and $\mu_r$ represent the relative permittivity and relative permeability, respectively.}

\section{MM model}
The model introduced by M. Mendoza and D. Muñoz in  \cite{MendozaMunoz}, henceforth referred to as the MM model, utilizes a local basis consisting of electric vectors $\Vec{e}^p_{ij}$ and velocity vectors $\Vec{v}^p_i$ within a cubic cell D3Q13. Additionally, magnetic vectors $\Vec{b}^p_{ij}$ are employed in a D3Q7 cell. The planes are labeled as $p = 1, 2, 3$, while $i = 1, 2, 3, 4$ represents the discretized velocity directions, with four directions per plane. Each velocity vector corresponds to two electric vectors denoted by $j = 0, 1$.

In this framework, the direction of the velocity vectors can be interpreted as the direction of energy flux, i.e., the pointing vector. Specifically, the velocity vectors are defined as follows:

\begin{equation}
\begin{array}{l}\vec{v}_{i}^{0}=\sqrt{2}(\cos ((2 i-1) \pi / 4), \sin ((2 i-1) \pi / 4), 0)\,, \\ \vec{v}_{i}^{1}=\sqrt{2}(\cos ((2 i-1) \pi / 4), 0, \sin ((2 i-1) \pi / 4))\,,\\
\vec{v}_{i}^{2}=\sqrt{2}(0, \cos ((2 i-1) \pi / 4), \sin ((2 i-1) \pi / 4))\,,\\
\vec{v}_0 = (0,0,0)\,.
\end{array}
\label{eq:VelocitySetMiller}
\end{equation}
The electric and magnetic vectors are given by
\begin{equation}
    \begin{array}{c}\vec{e}_{i 0}^{p}=\frac{1}{2} \vec{v}^{p}_{[(i+2) \quad \bmod 4]+1} \quad, \quad \vec{e}_{i 1}^{p}=\frac{1}{2} \vec{v}^{p}_{[i \bmod 4]+1} \\ \vec{b}_{i j}^{p}=\vec{v}_{i}^{p} \times \vec{e}_{i j}^{p} .\end{array}
    \label{eq:ElectromagneticVectorsMiller}
\end{equation}

These vector sets exhibit appropriate sum relations, as demonstrated in Equation (11) of reference \cite{MendozaMunoz}.

For the electric and magnetic fields, there exists a distribution function $f^{p(r)}_{ij}$ associated with each $p, i, j$. Here, $r = 0$ denotes the electric field, while $r = 1$ represents the magnetic field. Moreover, two distribution functions are assigned to the rest direction for both electric and magnetic fields. Consequently, the total number of distribution functions amounts to $2 \times 2 \times 3 \times 4 + 2 = 50$.

The macroscopic fields are derived from the distribution functions through the following equations:

\begin{equation}
\begin{array}{c}\epsilon_r \vec{E}^{\prime} = \vec{D}^{\prime}=\sum_{i=1}^{4} \sum_{p=0}^{2} \sum_{j=0}^{1} f_{i j}^{p(0)} \vec{e}_{i j}^{p} \\ \vec{B}=\sum_{i=1}^{4} \sum_{p=0}^{2} \sum_{j=0}^{1} f_{i j}^{p(1)} \vec{b}_{i j}^{p} \\ \rho_{c}=f_{0}^{(0)}+\sum_{i=1}^{4} \sum_{p=0}^{2} \sum_{j=0}^{1} f_{i j}^{p(0)}\end{array}
\label{eq:MacroscopicFieldsMiller}
\end{equation}
It is remarkable to say that the displacement field  $\vec{D}^{\prime}$ is an auxiliary field. The real field which satisfies Maxwell's equations is 
\begin{equation}
   \frac{\vec{D}}{\epsilon_r} = \vec{E}  =\vec{E}^{\prime}-\frac{\mu_{0}}{4 \epsilon_{r}} \overrightarrow{J}\,, 
   \label{eq:MacroscopicEMiller}
\end{equation}
where $\vec{J}$ is the current source in Maxwell's equation. In particular, for ohmic elements $\vec{J} = \sigma \vec{E}$, that is, 
\begin{equation}
    \vec{J}=\frac{\sigma}{1+\frac{\mu_{0} \sigma}{4 \epsilon_{r}}} \frac{\vec{D}^{\prime}}{\epsilon_r}\,.
    \label{eq:MacorscopicCurrentMiller}
\end{equation}
Furthermore, the $H-$field is given by
\begin{equation}
    \vec{H}=\frac{\vec{B}}{\mu_{r}}\,.
    \label{eq:MacroscopicBMiller}
\end{equation}

To avoid dissipative effects, the dynamics of the Boltzmann equation adopt the BGK scheme with a collision time $\tau = 1/2$.

The equilibrium functions are defined as follows:
\begin{equation}
\begin{aligned}
     f_{i j}^{p(0) \mathrm{eq}}(\vec{x}, t)&=\frac{1}{16} \vec{v}_{i}^{p} \cdot \vec{J}^{}+\frac{\epsilon_r}{4} \vec{E}^{} \cdot \vec{e}_{i j}^{p}+\frac{1}{8 \mu_r} \vec{B} \cdot \vec{b}_{i j}^{p}\,,\\
     f_{i j}^{p(1) \mathrm{eq}}(\vec{x}, t)&=\frac{1}{16} \vec{v}_{i}^{p} \cdot \vec{J}^{}+\frac{1}{4} \vec{E}^{} \cdot \vec{e}_{i j}^{p}+\frac{1}{8} \vec{B} \cdot \vec{b}_{i j}^{p}\,,\\
     f_{0}^{(0) \mathrm{eq}}(\vec{x}, t)&=f_{0}^{(1) \mathrm{eq}}(\vec{x}, t)=\rho_{c}\,.
     \label{eq:equilibriumsMiller}
\end{aligned}
\end{equation}
The advection collision terms are applied conventionally. The BGK collision step:
\begin{equation}
    \begin{aligned}
        f^{p(r)\prime}_{i j}(\vec{x}, t)&=2 f_{i j}^{p(r) e q}(\vec{x}, t)-f_{i j}^{p(r)}(\vec{x}, t)\,,\\
        f_0^{0\prime}(\vec{x}, t)&=2 f_{0}^{0 e q}(\vec{x}, t)-f_{0}^{0}(\vec{x}, t)\,.
        \label{eq:CollisionMiller}
    \end{aligned}
\end{equation}
The advection step:
\begin{equation}
    \begin{aligned}
    f_{0}^{0}\left(\vec{x}, t+\Delta t\right)&=f^{0\prime}_{0}(\vec{x}, t)\,,\\
        f_{i j}^{p(r)}\left(\vec{x}+\vec{v}_{i} \Delta t, t+\Delta t\right)&=f^{p(r)\prime}_{i j}(\vec{x}, t)\,.
    \end{aligned}
\end{equation}
Interpreting the physical meaning of the equilibrium functions is not straightforward. However, M. Mendoza and D. Muñoz claim that these functions can be regarded as perturbations in the energy density.
\subsection{Maxwell's equations obtained by the MM model}
Finally, the MM model successfully reproduces the Maxwell equations for linear non-dispersive media. Its proof is based on the Chapman-Enskog expansion in \cite{MendozaMunoz}.
\begin{equation}
    \begin{array}{c}\frac{\partial \rho_{c}}{\partial t}+\nabla \cdot \vec{J}^{}=0 \,,\\ \nabla \times \vec{E}^{}=-\frac{\partial \vec{B}}{\partial t} \,,\\ \nabla \times \vec{H}= \vec{J}^{}+ \frac{\partial \vec{D}^{}}{\partial t}.\end{array}
\end{equation}
to ensure compliance with Gauss's laws, the equations in the MM model must be satisfied at time $t=0$. In the case of vacuum, the speed of electromagnetic waves in the MM model is equal to $1/\sqrt{2}$ in automaton units, where $\epsilon_0 = 1$ and $\mu_0 = 2$. It is worth noting that exceeding this speed limit can lead to numerical instabilities.

The authors of the model explain that this phenomenon is attributed to the CFL (Courant-Friedrichs-Lewy) condition. The CFL condition sets a constraint on the time step in numerical simulations to maintain stability.
\section{HV model}
The model proposed by A. Hauser and L. Verhey in \cite{HV2017}, referred to as the HV model, was introduced based on the model proposed by Y. Liun and G. Yan in \cite{LiuYan}. According to the authors, unlike the MM model, the HV model remains stable even in the presence of non-smooth transitions at interfaces between media with different permeability and permittivity.

The discretized set of velocity directions is represented by a D3Q7 cubic cell, with velocity vectors $\Vec{v}_i$ for $i = 1,...,6$, and $\Vec{v}_0$ as the rest vector. Similarly, for each $i\geq 1$, there is a pair of electromagnetic vectors $\Vec{e}_i$ and $\Vec{b}_i$. Furthermore, for each electromagnetic vector, there are three scalar distribution functions associated with the $x$, $y$, and $z$ components, resulting in the representation of the distribution function as 3D vectors.

The velocity vectors can be expressed as:
\begin{equation}
\begin{aligned}
\Vec{v}_1 &= (1,0,0), & \Vec{v}_2 &= (0,1,0), & \Vec{v}_3 &= (-1,0,0),\\ 
\Vec{v}_4 &= (0,-1,0), & \Vec{v}_5 &= (0,0,-1), & \Vec{v}_6 &= (0,0,1).
\end{aligned}
\end{equation}
In total, there are $2 \times 6 \times 3 = 36$ scalar distribution functions or 12 vector distribution functions.

In the original paper \cite{HV2017}, there is an error in equation (B1b). The correct equation should be:

\begin{equation}
\sum_{i=1}^{6} \Vec{v}_{\alpha, i} \cdot \Vec{v}_{\beta, i} = 2 \delta_{\alpha \beta}.
\label{eq:Correct}
\end{equation}

This error impacts the computations in the paper, including the distribution function. Therefore, considering the correction, the equilibrium distributions are given by:

\begin{equation}
    \begin{aligned}
\Vec{e}_i^{\mathrm{eq}} &= \frac{1}{6}\left( \Vec{D} - 3 \Vec{v}_i \times \frac{\Vec{B}}{\mu}\right), \\
\Vec{h}_i^{\mathrm{eq}} &= \frac{1}{6}\left(\Vec{B} + 3 \Vec{v}_i \times \frac{\Vec{D}}{\epsilon}\right).
\end{aligned}
\label{eq:equilibriumHV}
\end{equation}
Note that the equilibrium functions bear similarities to Maxwell's equations when considering $\nabla$ as a vector that is a scalar multiple of $\Vec{v}_i$. This notion is particularly relevant in the context of harmonic plane waves, where $\nabla$ is parallel to the wave vector (and pointing vector). This observation suggests that the equilibrium functions can be interpreted as perturbations of the Maxwell's equations themselves.

Similar to the MM model, we use $r=0,1$ to denote the electric and magnetic vectors, respectively, such that $\Vec{f}_i^{(0)} = \Vec{e}_i$ and $\Vec{f}_i^{(1)} = \Vec{b}_i$.

The macroscopic fields can be obtained from the distribution functions as follows:

\begin{equation}
\begin{aligned}
\epsilon \Vec{E} = \Vec{D}(\boldsymbol{r}, t) &= \sum_{i=1}^{6} \Vec{e}_i(\boldsymbol{r}, t), \\
\mu \Vec{H} = \Vec{B}(\boldsymbol{r}, t) &= \sum_{i=1}^{6} \Vec{h}_i(\boldsymbol{r}, t).
\end{aligned}
\label{eq:MacroscopicFieldsHV}
\end{equation}

Similarly to the MM model, the relaxation time $\tau = 1/2$ is used, and the BGK collision step is applied:

\begin{equation}
\begin{aligned}
\Vec{f}_i^{(r)'}(\Vec{x}, t) &= 2 \Vec{f}_i^{(r)e q}(\Vec{x}, t) - \Vec{f}_i^{(r)}(\Vec{x}, t).
\end{aligned}
\label{eq:CollisionHV}
\end{equation}

The advection step is given by:

\begin{equation}
\begin{aligned}
\Vec{f}_i^{(r)}(\Vec{x}+\Vec{v}_{i} \Delta t, t+\Delta t) &= \Vec{f}_i^{(r)'}(\Vec{x}, t).
\end{aligned}
\end{equation}

\subsection{Maxwell's equations obtained by the HV model}

As demonstrated in \cite{HV2017}, the equations obtained through the Chapman-Enskog expansion are:

\begin{equation}
    \begin{array}{c}
    \nabla \times \vec{E}=-\frac{\partial \vec{B}}{\partial t}\,, \\
    \nabla \times \vec{H}= \frac{\partial \vec{D}}{\partial t}
    \end{array}
\end{equation}

Following the proposal in \cite{Hanasoge_2011,ZhaoliZheng}, we introduce a source term $\Vec{T}_i$ for each $i$ in the electric distribution function ($r = 0$), defined as:
\begin{equation}
    \Vec{T}_i = -\frac{1}{2}\left(  \Vec{J} \cdot \Vec{v}_i \right) \Vec{v}_i,
    \label{eq:SourceTermHV}
\end{equation}
then the equation \eqref{eq:CollisionHV} is modified for $r=0$ as:
\begin{equation}
\begin{aligned}
\Vec{f}_i^{(0)'}(\Vec{x}, t) &= 2 \Vec{f}_i^{(0)e q}(\Vec{x}, t) - \Vec{f}_i^{(0)}(\Vec{x}, t) + \Vec{T}_i.
\end{aligned}
\label{eq:CollisionmodifiedHV}
\end{equation}
Thus, we obtain the Maxwell's equations:
\begin{equation}
    \begin{array}{c}
    \nabla \times \vec{E}=-\frac{\partial \vec{B}}{\partial t}\,, \\
    \nabla \times \vec{H}= \Vec{J} + \frac{\partial \vec{D}}{\partial t}
    \end{array}
\end{equation}
The HV model with the additional source term is referred to as the modified HV model.

In the HV model, the speed of electromagnetic waves in the vacuum is equal to $1/3$ in lattice units. Values greater than this limit can result in numerical instabilities. However, the authors of the model do not provide an explanation for this specific limit.
\section{Numerical test and comparison}

\subsection{Gaussian Pulse Crossing Dielectric Interface}
For the first comparison, we simulated a Gaussian pulse of the form
\begin{equation}
    \begin{aligned}
        \Vec{B} &= E_0/C \exp(-(z-z_0)^2/(2\alpha^2)) \, \hat{y}\,.\\
        \Vec{E} &= E_0 \exp(-(z-z_0)^2/(2\alpha^2)) \, \hat{x}\,,
    \end{aligned}
    \label{eq:Initialpulse}
\end{equation}
where $\alpha = 0.05\cdot L_z/\sqrt{2}$ and $z_0 = L_z/2 - L_z/6$ .

The pulse travels from the vacuum $\epsilon_{1,r} = 1$ to a media with $\epsilon_{2,r} = \epsilon_{r}=2.5$ and $\mu_{r}=1$. According to the theory, the reflected and transmitted amplitude are such that
\begin{gather}
    \frac{A_{ref}}{A_{inc}} = \frac{\sqrt{r}-1}{\sqrt{r}+1} \,,\\
    \frac{A_{trans}}{A_{inc}} = \frac{2}{\sqrt{r}+1}\,,
\end{gather}
being $r = \epsilon_{2,r}/\epsilon_{1,r}$ and,  $A_{ref}$ and $A_{trans}$ the amplitude of the reflected and transmitted pulse, respectively \cite{jackson_classical_1999}.


\begin{figure}[t]
\centering
\includegraphics[width=0.55\textwidth]{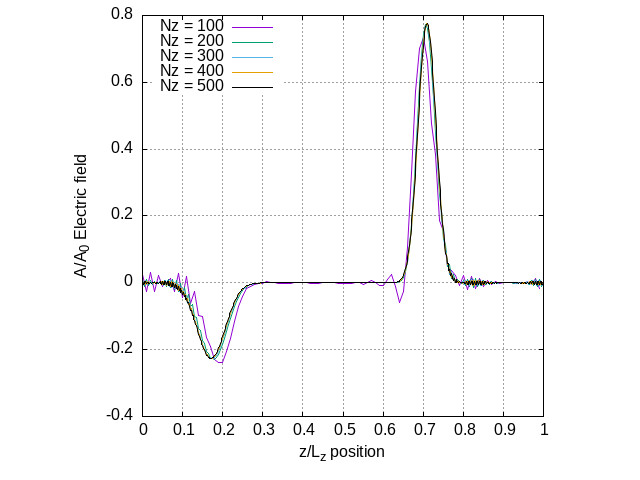}
\caption{Normalized electric amplitudes $A/A_0$ of a Gaussian pulse crossing a dielectric interface simulated using the MM model for different grid refinements labeled by the number of cells $N_z$. The region where $z/L_z > 0.5$ corresponds to a dielectric medium with $\epsilon_r = 2.5$, while the region where $z/Lz < 0.5$ corresponds to the vacuum ($\epsilon_r = 1.0$). The reflected and transmitted pulse can be observed.}
\label{fig:MM_refinement} 
\end{figure}



\begin{figure}[H]
\centering
\begin{subfigure}[b]{0.50\textwidth}
   \includegraphics[width=1\linewidth]{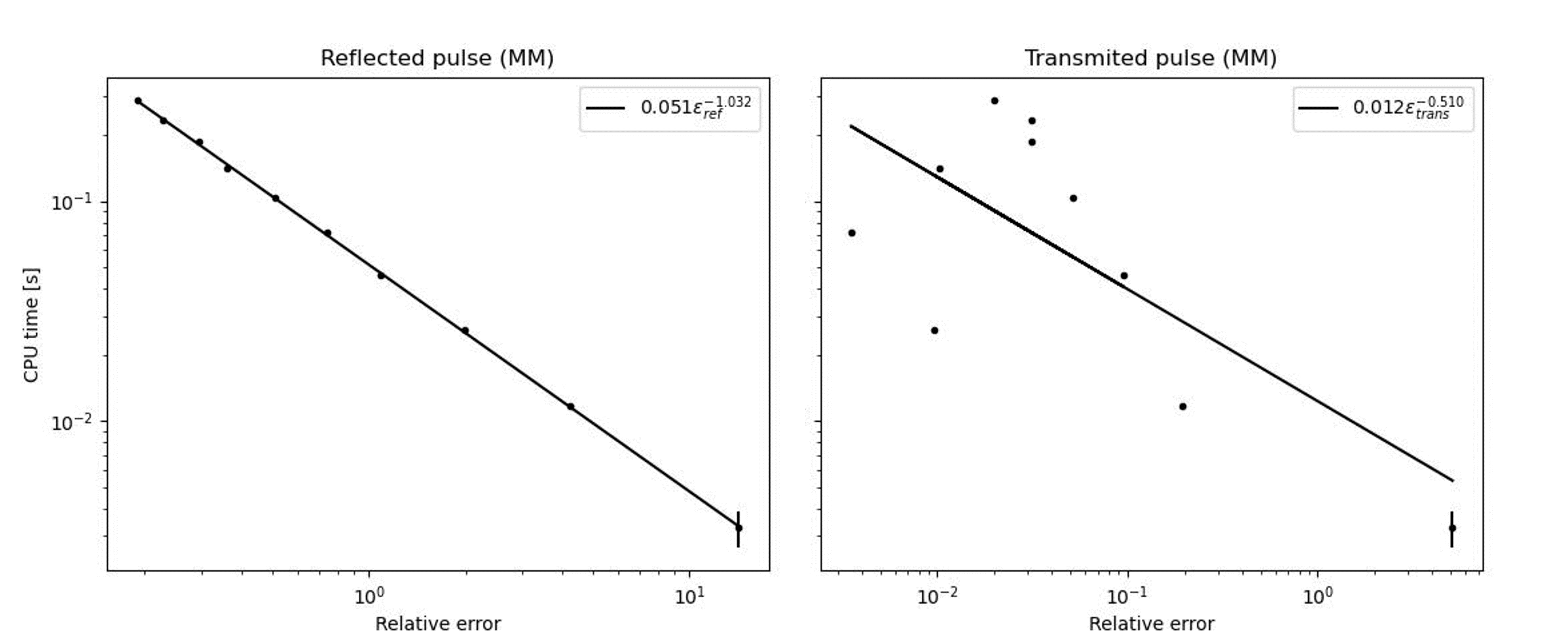}
   \caption{}
   \label{fig:pulse_MM} 
\end{subfigure}

\begin{subfigure}[b]{0.50\textwidth}
   \includegraphics[width=1\linewidth]{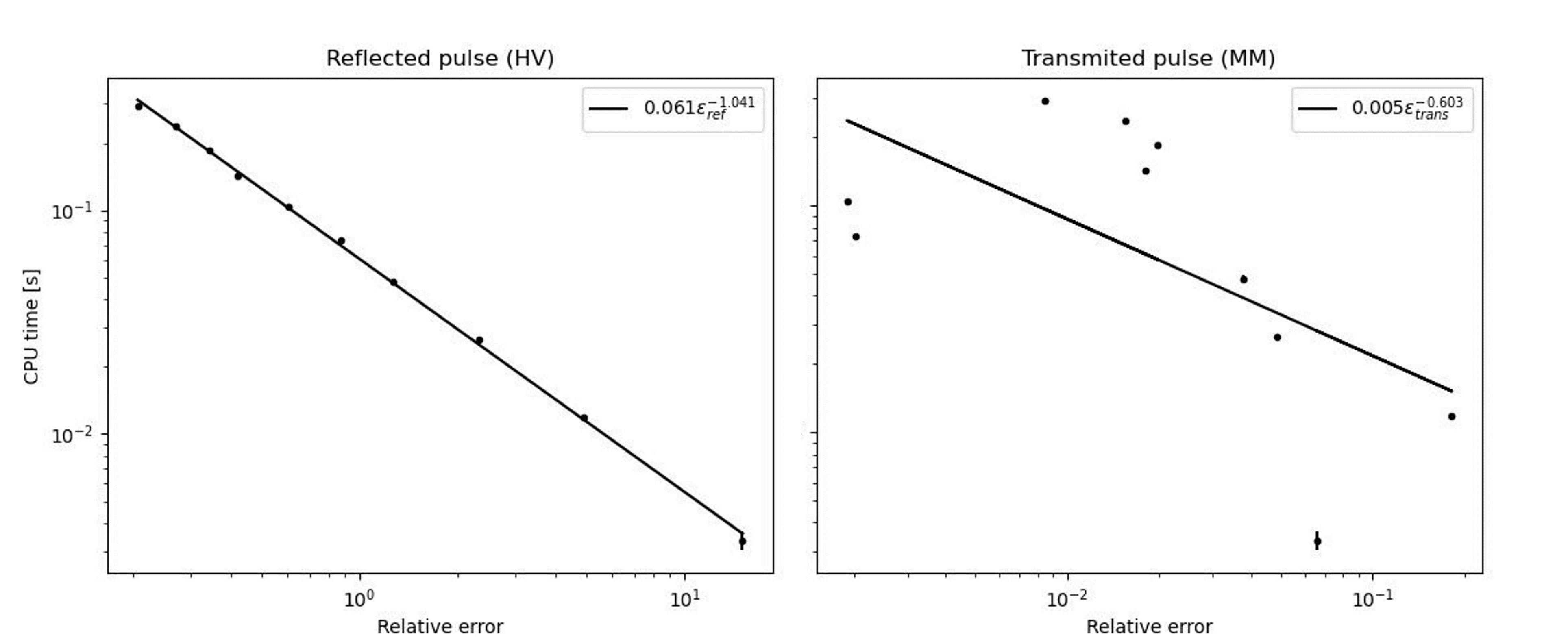}
   \caption{}
   \label{fig:pulse_HV}
\end{subfigure}
\caption{CPU time versus simulation error in the reflected electric amplitude for the dielectric pulse test.}
\label{fig:pulse_ref}
\end{figure}

Our simulations consisted of a one-dimensional grid, progressively refined while keeping the physical size unchanged. In figure  (\ref{fig:MM_refinement}), the shape of the transmitted and reflected pulse is depicted for different refinements using the MM model. For each grid, we measured the CPU time only for the integration of Maxwell's equations with the LBM, i.e., our measurement excluded the CPU time for the initialization of variables or printing. Similarly, we also gauged the relative error in the amplitude of the simulated reflected and transmitted pulse, obtaining the curves in figure (\ref{fig:pulse_ref}). Importantly, the time measurement was made several times for each refinement to compute an average and error bars.




The plots show that the MM scheme requires less CPU time for a given relative error. Remarkably, the relation CPU time-error for the transmitted pulse is erratic for both simulations. Furthermore, the error bars for both formulations are generally small, suggesting that this behavior is intrinsic in the two schemes.
\begin{figure}[H]
\centering
\begin{subfigure}[b]{0.4\textwidth}
   \includegraphics[width=1\linewidth]{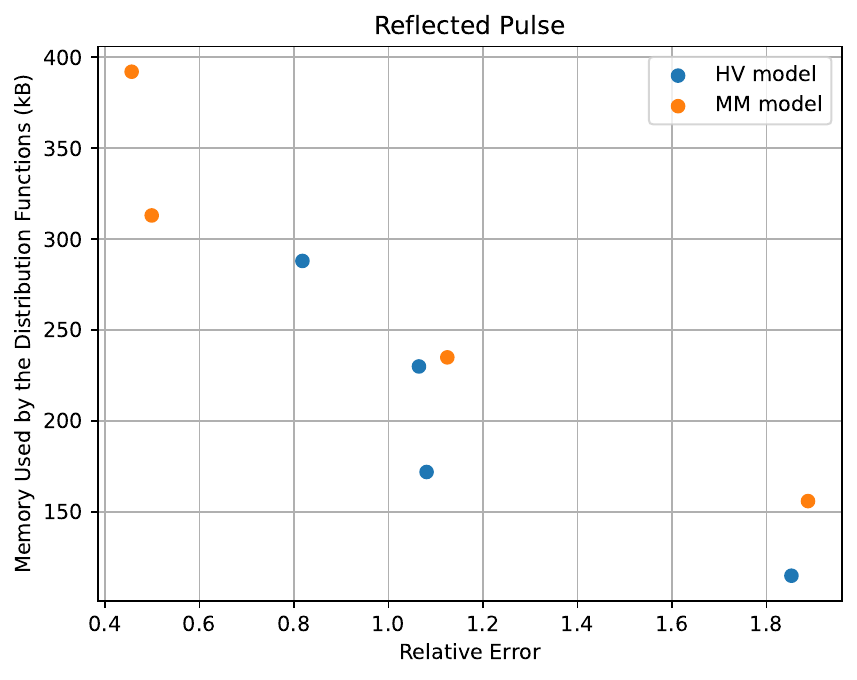}
   \caption{}
   \label{fig:ref_mem_trans} 
\end{subfigure}

\begin{subfigure}[b]{0.4\textwidth}
   \includegraphics[width=1\linewidth]{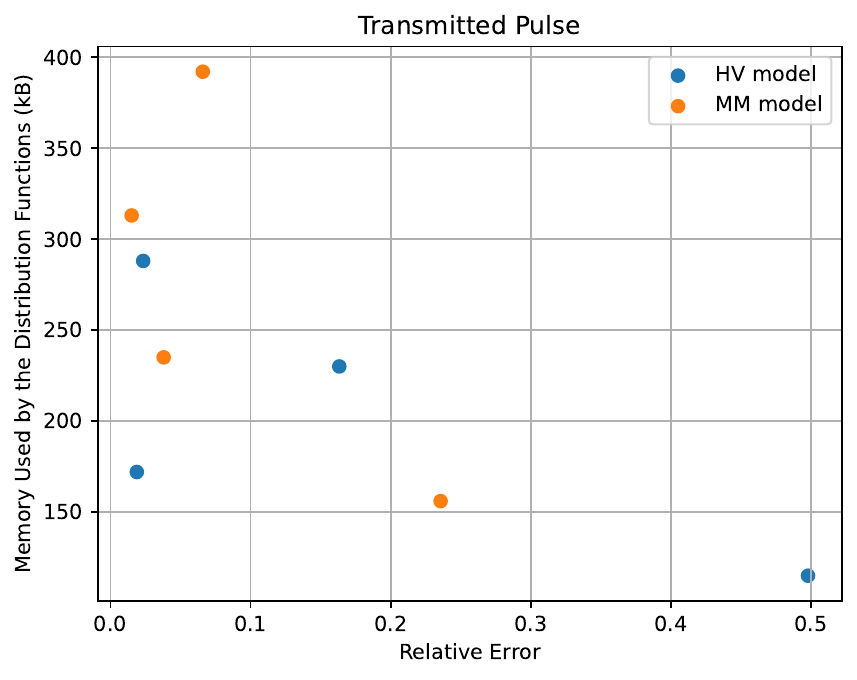}
   \caption{}
   \label{fig:ref_mem_trans}
\end{subfigure}
\caption{Memory usage of the distribution functions for both models as a function of the relative errors for the dielectric pulse test.}
\label{fig:pulse_mems} 
\end{figure}

Regarding the behavior of the relative error for different refinement levels, it was observed in both schemes that the error decreases following a power law as the grid refinement is increased. However, it is notable that the MM model exhibits the most significant reduction in numerical errors with the implementation of grid refinement.

A quick measurement of the memory used by the distribution functions was also conducted as a function of the relative errors (figure \ref{fig:pulse_mems}). The memory requirement was observed to increase for lower relative errors in both schemes. However, the MM model exhibits a higher memory requirement than the HV model due to the difference in the number of distribution functions, as mentioned in Sections II and III.
\subsection{Skin Effect}
The Skin effect describes the exponential decay of a plane wave's amplitude after penetrating a conducting material. Theoretically, the amplitude of the electric field inside the conductor is given by the expression: 

\begin{equation}
    A_{\text{Theo}} = A_0 \exp(-z/\delta) 
\end{equation}

where $A_0$ represents the amplitude outside of the conductor, and $\delta$ denotes the skin thickness, which can be expressed as \cite[p. 130]{jordan1995electromagnetic}:

\begin{equation}
    \delta = \sqrt{\frac{2}{\sigma \mu \omega}} \sqrt{\sqrt{1+\left( \frac{\omega \epsilon}{\sigma}\right)^2} + \frac{\omega \epsilon}{\sigma}} \,.
\end{equation} 

where $\omega$ corresponds to the angular frequency of the wave. 


\begin{figure}[t]
\centering
\includegraphics[width=0.55\textwidth]{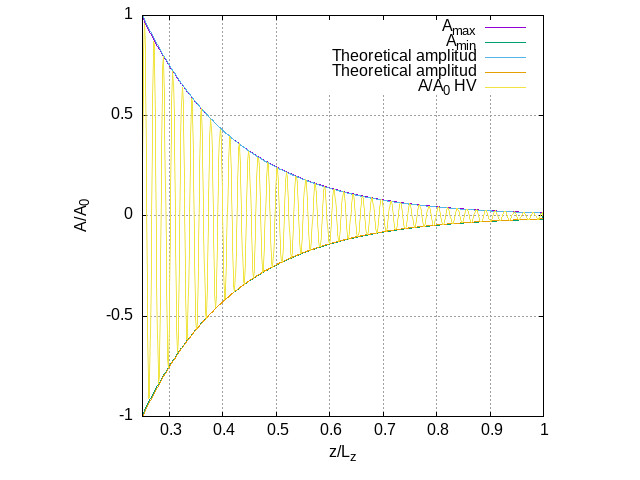}
\caption{Normalized electric field amplitude inside a conductor simulated with the HV model. A plane wave encounters a conductive medium at $z/L_z \geq 0.25$ with conductivity $\sigma = 0.1 \cdot \epsilon/T $ .}
\label{fig:HV-Skin} 
\end{figure}
We simulate a plane wave in one dimension, imposing the electromagnetic field at $z = 0$ with a wavevector to the right.
\begin{equation}
    \begin{aligned}
        \Vec{B} &= E_0/C \sin (\omega t) \hat{y} \,\\
        \Vec{E} &= E_0 \sin (\omega t) \hat{x} \,,
    \end{aligned}
\end{equation}
where $\omega = 2\pi/T$, and the period is $T = 17.68 \cdot 10^{-3} L_z/C$. A conductive medium is placed at $z/L_z \geq 0.25$ with conductivity $\sigma = 0.1 \cdot \epsilon / T$ .

The behavior of the oscillating wave and its amplitude after penetrating the conductor is presented in the figure (\ref{fig:HV-Skin}) for the HV model.

\begin{figure}
\centering
\begin{subfigure}[b]{0.5\textwidth}
   \includegraphics[width=1\linewidth]{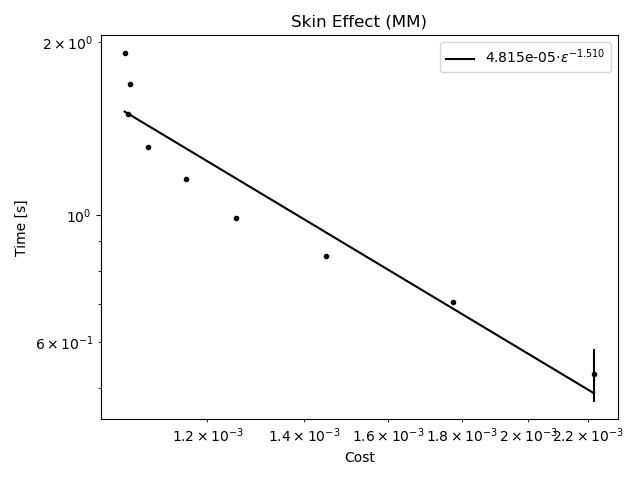}
   \caption{}
   \label{fig:SE_MM} 
\end{subfigure}

\begin{subfigure}[b]{0.5\textwidth}
   \includegraphics[width=1\linewidth]{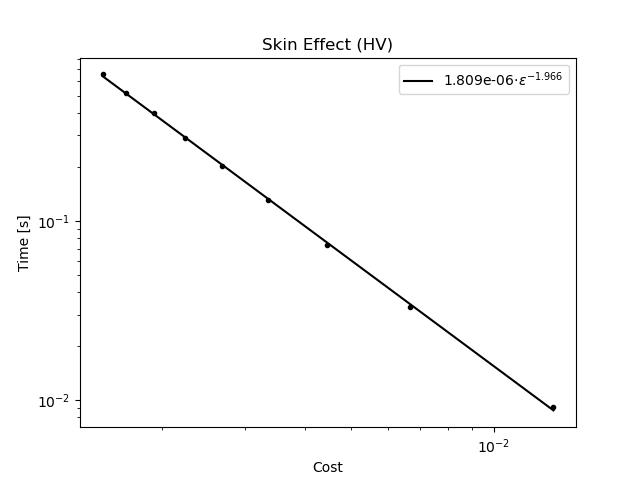}
   \caption{}
   \label{fig:SE_HV}
\end{subfigure}
\caption{CPU time versus cost for the Skin effect.}
\label{fig:SE}
\end{figure}

We proceeded similarly to the Gaussian pulse to compare the errors and computing time. We utilized a one-dimensional grid with several refinements and measured the time and error for each grid, computing an average and error for the time. The only difference lies in the error measurement, which was modeled with a cost function
\begin{equation}
\label{eq:cost}
    C = \sum_{j}\bigg(\frac{A_{\text{sim}}-A_{\text{theo}}}{A_{\text{theo}}}\bigg)^{2}
\end{equation}
where $A_{\text{sim}}$ and $A_{\text{theo}}$ are the simulated and theoretical electric field amplitudes after penetrating the conductor, and the sum is carried out over the cells occupied by the conductor. As figure (\ref{fig:SE}) shows, where both time measurements overlap, the MM scheme produces a smaller cost; additionally, the polynomial fit shows that for ever smaller costs, the MM model requires less computation time.

\subsection{Electric Dipole Radiation}
An oscillating electric or magnetic dipole represents one of the simplest antenna systems available and is extensively treated in many textbooks \cite{jackson_classical_1999}, the reason why is an ideal example to test performance for simulation of radiating systems. 

The simulation involves a Hertz dipole with specific parameters. Instead of a point source, a localized density current associated with a dipole is simulated. The current density is given by 
\begin{equation}
    J = J_0 \exp\left(-\alpha \left( \vec{x} - \vec{x_d}\right)^2\right) \sin(\omega t) \hat{z}.
\end{equation}
where \(\alpha = 0.5\), the amplitude of the current density \(J_0\) is set to 0.0001 and $\vec{x_d}$ are the dipole coordinates. The period \(T\) is calculated as \(\frac{17.68}{100.0} \frac{Lz}{C}\), and \(\omega\) is set to \(\frac{2 \pi}{T}\). The associated amplitude of the dipole is \cite{MendozaMunoz}\cite[Chapter 9]{jackson_classical_1999}
\begin{equation}
    p = \frac{J_0}{\omega}\left(\frac{\pi}{\alpha}\right)^{1.5} \,.
\end{equation}
The vacuum impedance \(Z_0\) is calculated as \(\sqrt{\frac{\mu_0}{\epsilon_0}}\), and the wavelength \(\lambda\) is determined as \(c \cdot T\), being $c$ the speed of light. The maximum simulated time is \(t_{\text{max}} = T \cdot\frac{70}{25.0}\) in the time test comparisons.

To obtain the radiation patterns, measurements are taken at a radius $R = \lambda \cdot n$, where $n = \frac{Lz}{2\lambda} - 2$ (two wavelengths less before reaching the boundary of the lattice domain). We use $t_{\text{max}} = \frac{R+2\lambda}{C}$, and measure the maximum energy flux during a whole period in the time interval $R + \lambda \leq t \leq R+2\lambda$.\\
The normalized theoretical radiation pattern in spherical coordinates $(\phi,\theta)$, taking the z-axis in the direction of the dipole, is $\sin^2(\theta)$, see \cite[Chapter 9]{jackson_classical_1999}. In figure (\ref{fig:HV-Dipole-Power}) is plotted the theoretical and simulated radiation pattern in a plane of constant $\phi$.

Finally, we gauged the same cost function (\ref{eq:cost}) and CPU time for the error and time measurement. Specifically, we compared the simulated and theoretical electric and magnetic field amplitudes along a line perpendicular to the dipole moment. From figure (\ref{fig:DIP}) is clear that in the studied interval, both models had very similar performance. Even more remarkable, for a given computation time, the electric and magnetic field costs are not the same for a given LB; furthermore, in the limit of small costs, MM performs better for the magnetic field but worse for the electric field.
\begin{figure}[H]
    \centering
    \begin{subfigure}[b]{0.55\textwidth}
        \centering
        \includegraphics[width=\textwidth]{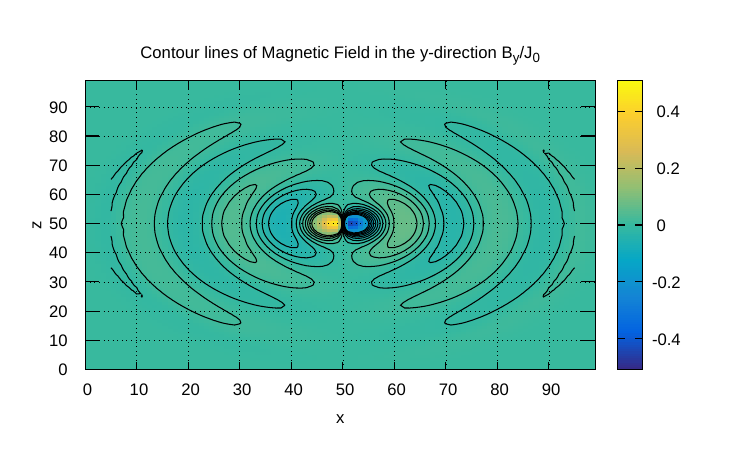}
        \caption{Contour lines of magnetic field $B_y/J_0$ in the x-z plane simulated by the MM model in automaton units.}
        \label{fig:MM-Dipole-Contour}
    \end{subfigure}
    \hfill
    \begin{subfigure}[b]{0.4\textwidth}
        \centering
        \includegraphics[scale=0.7]{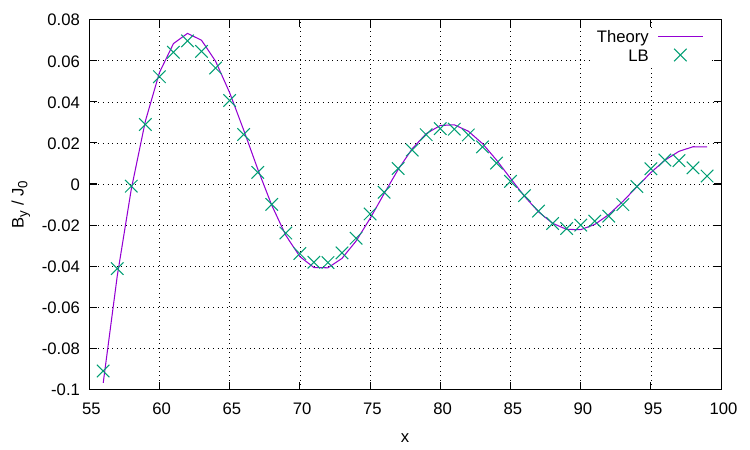}
        \caption{Theoretical and simulated magnetic field along the x-axis using the MM model in automaton units.}
        \label{fig:MM-Dipole-By-axisx}
    \end{subfigure}
    \caption{Comparison of magnetic field simulations using the MM model.}
    \label{fig:MM-Dipole}
\end{figure}

\begin{figure}[H]
\centering
\includegraphics[width=0.5\textwidth]{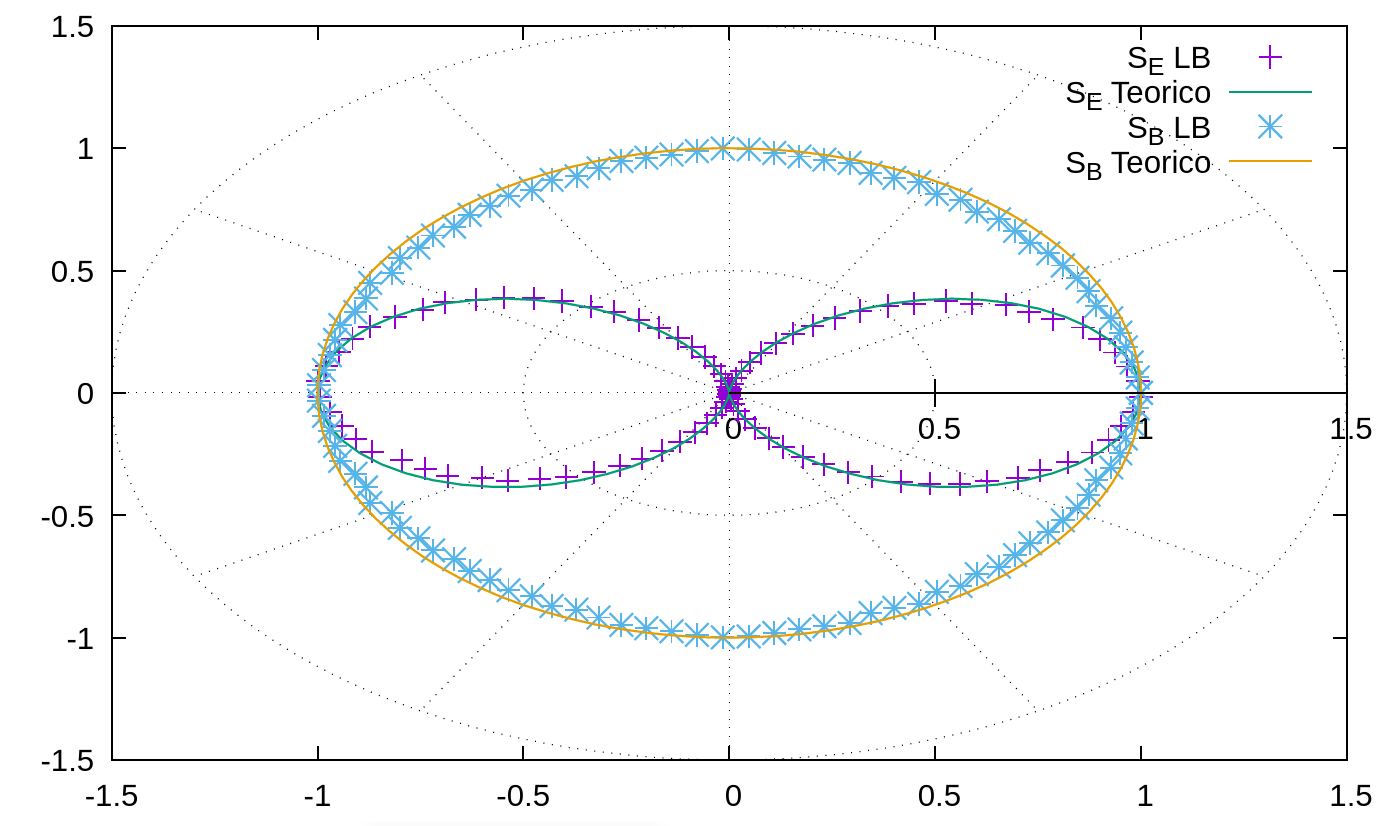}
\caption{Normalized theoretical and simulated slice of the radiation pattern of the Hertz dipole using the HV model.}
\label{fig:HV-Dipole-Power} 
\end{figure}
\begin{figure}[H]
\centering
\begin{subfigure}[b]{0.5\textwidth}
   \includegraphics[width=1\linewidth]{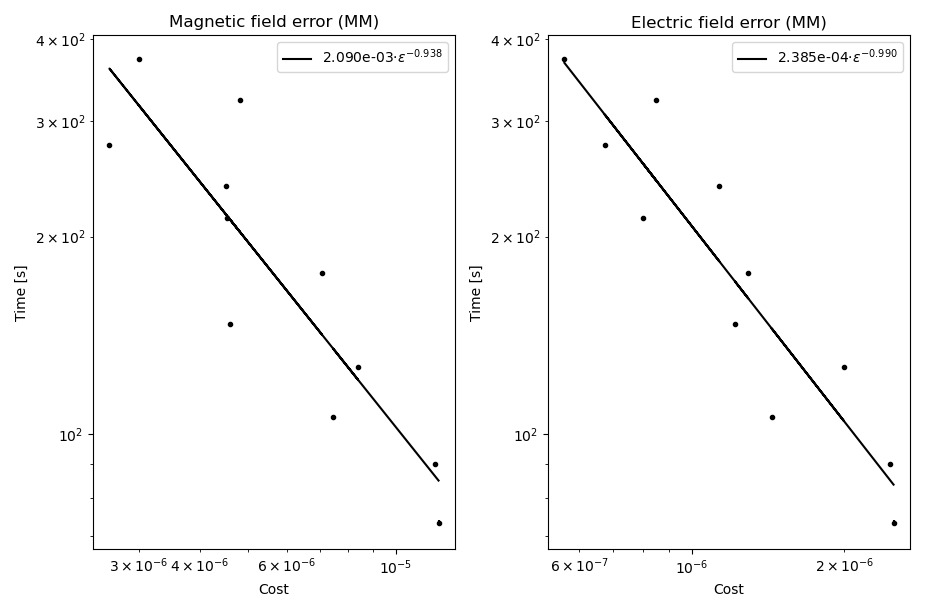}
   \caption{}
   \label{fig:DIP_MM} 
\end{subfigure}

\begin{subfigure}[b]{0.5\textwidth}
   \includegraphics[width=1\linewidth]{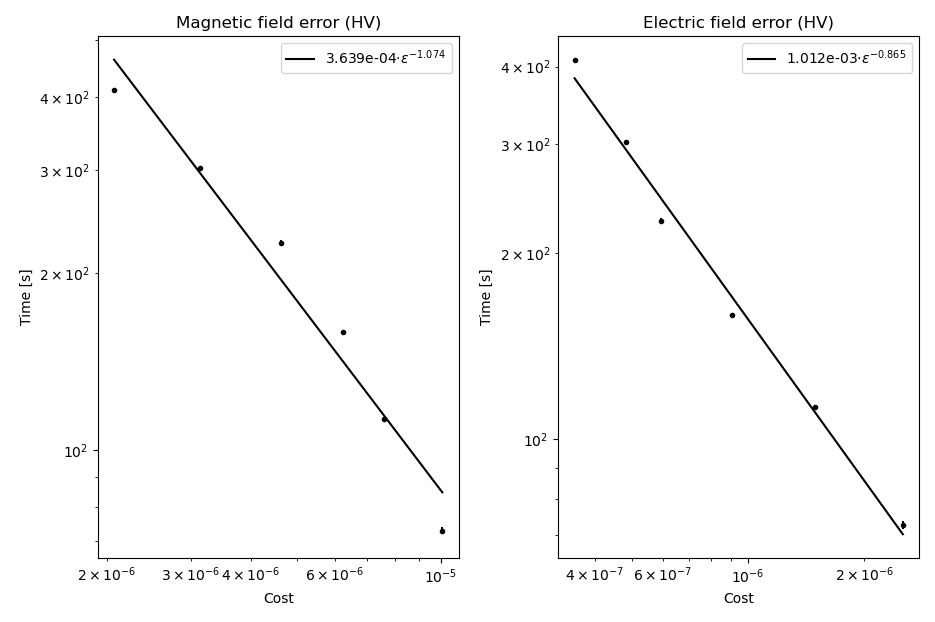}
   \caption{}
   \label{fig:DIP_HV}
\end{subfigure}
\caption{CPU time versus cost for the radiating dipole.}
\label{fig:DIP}
\end{figure}
\subsection{Gaussian Pulse Crossing Non-uniform media}
The setup was similar to the section on the dielectric pulse crossing one interface. However, in this case, we have four different mediums with relative permitivities of $1, 1.3, 2, 3$. The corresponding interfaces are located at $z_0, z_0 + d, z_0 + 3d$, respectively, where $z_0 = \frac{L_z}{2}$ and $d = \frac{L_z}{20}$.

The simulation space consisted of $L_z$ cells, and the pulse was positioned at $z_0 - \frac{L_z}{30}$ using an $\alpha = \frac{L_z}{100\sqrt{2}}$ in equation \eqref{eq:Initialpulse}. To achieve an error of 0.03 $\%$ between the simulated and theoretical transmitted pulses using both methods, the MM model requires 18 seconds of CPU time and 4000 cells, which corresponds to 200,000 distribution functions. On the other hand, the HV model only requires 10 seconds of CPU time and 3000 cells, totaling 108,000 scalar distribution functions. Figures 6 and 7 depict the pulse after crossing all the interfaces.

\begin{figure}[H]
\centering
\includegraphics[width=0.55\textwidth]{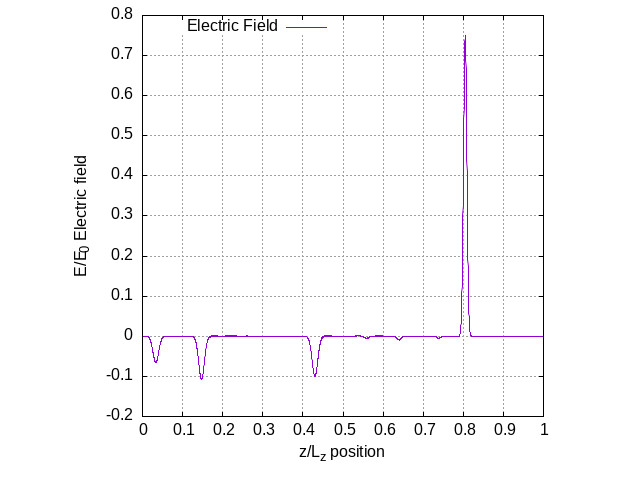}
\caption{Dielectric pulse crossing several interfaces using the HV model.}
\label{fig:HV-MInterface-Pulse} 
\end{figure}

\begin{figure}[H]
\centering
\includegraphics[width=0.55\textwidth]{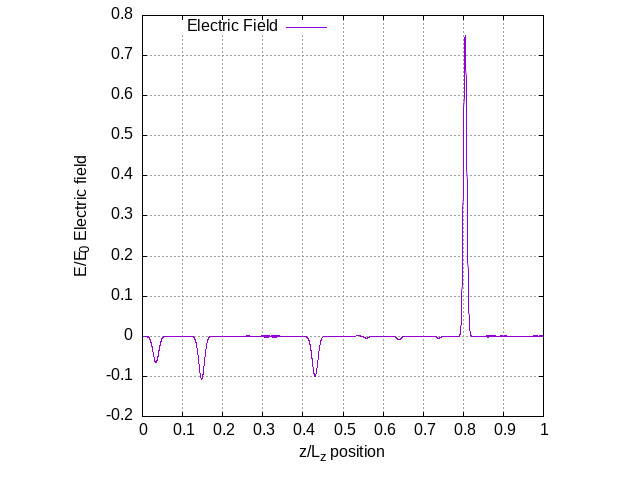}
\caption{Dielectric pulse crossing several interfaces using the MM model.}
\label{fig:MM-MInterface-Pulse} 
\end{figure}

\section{Conclusion}

Both models are suitable alternatives for simulating electrodynamics phenomena. They exhibited similar behavior in terms of computation time and error. However, the HV model outperformed the MM model in a test involving a Gaussian pulse crossing a non-uniform medium. Despite this, the MM model generally performed slightly better than the HV model in most tests. Nevertheless, the HV model demonstrated lower memory consumption due to its reduced number of distribution functions. Therefore, the HV model may be preferred as it requires less memory, while yielding similar errors to the MM model with equal CPU times. It appears that reducing the number of distribution functions does not significantly affect overall performance but improves memory requirements. The trade-off of the HV model is that it requires more iterations because the speed of light in automaton units is lower than that of the MM model. Future works could explore more LBMs and further test the validity of this claim.

\section*{Authorship contribution statement}

JR implemented the Lattice-Boltzmann algorithms and proposed the modified HV model. AS and JH performed the performance tests. AS and JR conceived the project. All three authors wrote and revised the article.

\section*{Acknowledgment}

The authors would like to express their gratitude to Dr. José Daniel Muñoz for his valuable comments and insights throughout the development of this research. Additionally, they would like to thank Dr. Rafael Rey for suggesting the dielectric pulse test conducted at multiple interfaces and to thank Dr.Andreas Hauser for clarifying some of our doubts. Their contributions greatly enhanced the quality and depth of this study. In addition, ChatGPT and Grammarly were utilized for writing style grammar correction.


%





\ifCLASSOPTIONcaptionsoff
  \newpage
\fi
\section*{APPENDIX A: Source term in the HV model}
\label{ap:appendixA}
\begin{equation}
    \sum_{i}\left(\partial_{t} f_{i}^{\mathrm{eq}}+\sum_{\alpha} v_{\alpha_{i}} \partial_{\alpha} f_{i}^{\mathrm{eq}}\right) \approx \sum_{i} T_i \,.
\end{equation}
Following this equation for the equilibrium functions $\Vec{f}_i^{(0)}$ we obtain in the left-hand side
\begin{equation}
    \frac{\partial \vec{D}}{\partial t} - \nabla \times \vec{H}\,.
\end{equation}
On the other hand, we get in the right-hand side
\begin{equation}
    \begin{aligned}
         \sum_{i = 1}^6 \Vec{T}_{i,\alpha}  &= -\frac{1}{2} \sum_{i = 1}^6 \left(   \Vec{J} \cdot \Vec{v}_i  \right) \Vec{v}_{i,\alpha} \,,\\
        &= -\frac{1}{2} \sum_{i = 1}^6   \Vec{J}_{\beta} \Vec{v}_{i,\beta} \Vec{v}_{i,\alpha} \,,\\
        &= -\frac{1}{2} \Vec{J}_{\beta} \sum_{i = 1}^6 \Vec{v}_{i,\beta} \Vec{v}_{i,\alpha} \,,\\
        &= -\frac{1}{2} \Vec{J}_{\beta} (2 \delta_{\alpha, \beta}) \,,\\
        &= - \Vec{J}\,.
    \end{aligned}
\end{equation}
Thus the Ampere-Maxwell equation is obtained.
\begin{equation*}
    \begin{aligned}
        \frac{\partial \vec{D}}{\partial t} - \nabla \times \vec{H} = - \Vec{J}
    \end{aligned}
\end{equation*}

\section*{Appendix B: PSEUDO-CODE}
In the following pseudocode, we outline the steps involved in the Lattice Boltzmann Automata simulation for solving a specific problem using the MM and modified HV models. The pseudocode describes the main algorithmic steps involved in the simulation, including initialization, collision, advection, and analysis. The specific equations and considerations for different models are also highlighted.
\begin{algorithmic}[1]
\State At $t=0$, impose all fields $\Vec{B}, \Vec{E}, \Vec{J}$.
\State Initialize all the distribution functions as the equilibrium functions evaluated in $\Vec{B}, \Vec{E}$ and $ \Vec{J}$.
\State Impose fields in the cells, if required.
\For{$t = 1$ to $t_{\text{max}}$}
    \State // In the collision, compute the macroscopic fields by summing the distribution functions, i.e, using eqs. \eqref{eq:MacroscopicFieldsMiller} to \eqref{eq:MacroscopicBMiller} for MM model and \eqref{eq:MacroscopicFieldsHV} for HV model
    \State Collision
    \State Impose fields if required.
    \State Advection
\EndFor
\State Analyze and plot.
\end{algorithmic}

\textbf{Note:} Sometimes, for instance, in the skin effect and in the Hertz dipole, the current is imposed during the collision step. In the MM model, the current is computed with equation \eqref{eq:MacorscopicCurrentMiller} after computing $\Vec{D'}$ with equation \eqref{eq:MacroscopicFieldsMiller}, and then it is used in the equilibrium functions \eqref{eq:equilibriumsMiller} during the collision step as in equation \eqref{eq:CollisionMiller}. On the other hand, in the HV model, the current is computed (for example as $\Vec{J} = \sigma \Vec{E}$) and then is passed to the source term $T$ as in equation \eqref{eq:SourceTermHV}.



\vfill

\end{document}